\documentclass [a4paper,12pt]{article}

\begin{document}

\begin{center}
{\large \bf A Remark on A Remark on Neutrino Oscillations Observed
in KamLAND Experiment.}

\vskip 0.2in

Anatoly Kopylov\\ Institute of Nuclear Research of Russian Academy
of Sciences \\ 117312 Moscow, Prospect of 60th Anniversary of
October Revolution 7A
\end{center}

\begin{abstract}
It is shown that equal magnitudes of the transitions $\bar \nu_e
\to \bar \nu_{\mu}$ and $\bar \nu_e \to \bar \nu_{\tau}$ in the
disappearance of reactor $\bar \nu_e$ discovered in the KamLAND
experiment just follows at $\theta_{23}=\pi/4$ and $\theta_{13}=0$ 
from pure symmetry of $\nu_{\mu}$ and $\nu_{\tau}$ states relatively 
the mass states.
\end{abstract}

It was shown in \cite{1} by the expressions (1)-(23) that in the
disappearance of reactor antineutrinos, discovered in the KamLAND
experiment \cite{2} the transitions $\bar \nu_e \to \bar
\nu_{\mu}$ and $\bar \nu_e \to \bar \nu_{\tau}$ have equal
magnitudes. The PMNS mixing matrix has absolutely symmetrical lines
for $\nu_{\mu}$ and $\nu_{\tau}$ relatively the mass states at
$\theta_{23}=\pi/4$ and $\theta_{13}=0$. Good illustrations of
this are presented, for example, in \cite{3}. It means that by
superimposing the system with the difference of phases between mass 
states 1 and 2 accumulated in propagation of $\nu_e$ in terms of
the flavor states $\nu_e$, $\nu_{\mu}$ and $\nu_{\tau}$, we have
no preference of $\nu_{\mu}$ over $\nu_{\tau}$ and vice versa.
This proves that the magnitudes of the transitions $\bar \nu_e \to
\bar \nu_{\mu}$ and $\bar \nu_e \to \bar \nu_{\tau}$ are indeed
equal.

\end{document}